\documentclass[aps,prl,twocolumn,showpacs,
amssymb]{revtex4}
\usepackage{epsfig}
\usepackage{dcolumn}
\usepackage{bm}
\usepackage{amsmath,amssymb}

\begin{document}
\parskip 5pt
\def \be {\begin{equation}}
\def \ee {\end{equation}}
\def \beq {\begin{equation}}
\def \eeq {\end{equation}}
\def \bea {\begin{eqnarray}}
\def \eea {\end{eqnarray}}
\def \nn {\nonumber}
\def \p {\prime}
\def \M {\mathbb{M}}
\newcommand{\MI}{\mathbb{M}}
\newcommand{\RI}{\mathbb{R}}
\newcommand{\HI}{\mathbb{H}}
\newcommand{\CI}{\mathbb{C}}
\newcommand{\TI}{\mathbb{T}}
\newcommand{\EI}{\mathbb{E}}
\def\etal{{\it et al}.} \def\e{{\rm e}} \def\de{\delta}
\def\dd{{\rm d}} \def\ds{\dd s} \def\ep{\epsilon} \def\de{\delta}
\def\goesas{\mathop{\sim}\limits} \def\al{\alpha} \def\vph{\varphi}
\def\Z#1{_{\lower2pt\hbox{$\scriptstyle#1$}}}
\def\X#1{_{\lower2pt\hbox{$\scriptscriptstyle#1$}}}
\def\MM#1{{\cal M}^{#1}} \def\Vm{V(\varphi_0)}
\def\VF{V\Z F} \font\sevenrm=cmr7 \def\ns#1{_{\hbox{\sevenrm #1}}}
\def\Vmin{V\ns{min}} \def\sgn{\,\hbox{sgn}}
\def\coshf{\cosh\left(\chi(u-u\Z0)\right)}
\def\CQG#1{Class.\ Quantum Grav.\ {\bf#1}}
\def\frn#1#2{{\textstyle{#1\over#2}}}
\def\ApJ#1{Astrophys.\ J.\ {\bf#1}} \def\AsJ#1{Astron.\ J.\ {\bf#1}}
\def\frn#1#2{{\textstyle{#1\over#2}}}


\title{Dynamical relaxation of dark energy: A solution to early
inflation, late-time acceleration and the cosmological constant
problem}

\author{Ishwaree P. Neupane\footnote{Corresponding author:
ishwaree.neupane@cern.ch}} \affiliation{Department of Physics and
Astronomy, University of Canterbury, Private Bag 4800,
Christchurch, New Zealand} \affiliation{Central Department of
Physics, Tribhuvan University, Kirtipur, Kathmandu, Nepal}
\author{Benedict M.~N.~Carter}
\affiliation{Department of Physics and Astronomy, University of
Canterbury, Private Bag 4800, Christchurch, New Zealand}

\begin{abstract}

In recent years different explanations are provided for both an
inflation and a recent acceleration in the expansion of the
universe. In this Letter we show that a model of physical interest
is the modification of general relativity with a Gauss-Bonnet term
coupled to a dynamical scalar-field as predicted by certain
versions of string theory. This construction provides a model of
evolving dark energy that naturally explains a dynamical
relaxation of the vacuum energy (gravitationally repulsive
pressure) to a small value (exponentially close to zero) after a
sufficient number of e-folds. The model also leads to a small
deviation from the $w=-1$ prediction of non-evolving dark energy.

\end{abstract}



\maketitle


Inflation, or a period of accelerated expansion in the early
universe, and cosmic acceleration at late times as pure
gravitational dynamics are unusual within the context of general
relativity (GR). Einstein back in 1917 amended his General Theory
of Relativity with a cosmological constant $\Lambda$ to achieve a
stationary universe, but it was later realized that a positive
$\Lambda$, in a cosmological context, gives rise to an
accelerating expansion of the universe. In modern parlance, a
positive $\Lambda$, more generally, a vacuum energy, is called
``dark energy", which makes up about 70\% of the matter--energy
content of the present universe. The right explanation of dark
energy is the greatest challenge faced by the current generation
of physicists and cosmologists.

A pure cosmological constant, $\Lambda\sim ({10}^{-3}{eV})^4$, as
a source of dark energy does not seem much likely, as there is no
conceivable mechanism to explain why it is so tiny and also why
should it be comparable to the present matter
density~\cite{Weinberg:1988cp} (see the
reviews~\cite{Padma:2002}). It is very problematic for particle
physics if $\Lambda$ is to be interpreted as the vacuum energy. If
dark energy is something fundamental, then it is natural to
attribute it to one or more scalar fields~\cite{Peebles:1988}. It
is not only because dark energy associated with the dynamics of
scalar fields which are uniform in space provides a mechanism for
generating the observed density
perturbations~\cite{Mukhanov:1990me} and a negative pressure
sufficient to drive the accelerating
expansion~\cite{Linde:1981mu}, but there is another reason well
motivated by fundamental physics.

Gravity is attractive and thus curves spacetime whose dynamics is
set entirely by the spacetime curvature: the Riemann curvature
tensor. However, inflation (or cosmic acceleration) is repulsive:
this is caused not by gravity altering its sense, attraction to
repulsion, but due to an extra source, or vacuum energy, which
covers every point in space and exerts gravitationally smooth and
negative pressure. The fundamental scalar fields abundant in
higher dimensional theories of gravity and fields, such as
string/M theory, are such examples. Most of the scalars in string
theory are the structure moduli associated with the internal
geometry of space, which do not directly couple with the curvature
tensor. But a scalar field associated with the overall size and
the shape of the internal compactification manifold generically
couples with Riemann curvature tensor, even in four
dimensions~\cite{Narain92a}.

The question that naturally arises is: what is the most plausible
form of a four-dimensional gravitational action that offers a
resolution to the dilemma posed by the current cosmic
acceleration~\cite{Bennett03a}, or the dark energy problem, within
a natural theoretical framework? In recent years, this problem has
been addressed in hundreds of papers proposing various kinds of
modification of the energy-momentum tensor in the ``vacuum" (i.e.,
the source of a gravitationally repulsive pressure), e.g., phantom
field~\cite{Caldwell99a}, $f(R)$ gravity that adds terms
proportional to inverse powers of $R$~\cite{Capozziello02} or the
$R^2$ terms, or both these effects at once~\cite{Capozziello},
braneworld modifications of Einstein's GR~\cite{DGP-2000}, etc..
Other examples of recent interest are cosmological
compactifications of string/M theory on a hyperbolic
manifold~\cite{Townsend03a} and on twisted
spaces~\cite{Ish05-twist}.

A very desirable feature of a theory of scalar-tensor gravity is
quasi-linearity: the property that the highest derivatives of the
metric appear in the field equations only linearly, so as to make
the theory ghost free. Interestingly, differential geometry offers
a particular combination of the curvature squared terms with such
behavior, known as the Gauss-Bonnet (GB) integrand:
$${\cal G}\equiv  R^2-4 R_{\mu\nu} R^{\mu\nu} + R_{\mu\nu\rho\sigma}
R^{\mu\nu\rho\sigma}.$$ All versions of string theory in 10
dimensions (except type II) include this term as the leading order
$\alpha^\prime$ correction~\cite{Callan85a}. A unique property of
the string effective action is that the couplings are field
dependent, and thus in principle space-time dependent. The
effective action for the system may be taken to be
\begin{equation}\label{dilatonGB}
S= \int d^{d}{x}\sqrt{-g}\left[
\frac{R}{2\kappa^2}-\frac{\gamma}{2} (\nabla\sigma)^2- V(\sigma)+
f(\sigma) {\cal G}\right],
\end{equation}
where $\kappa$ is the inverse Planck mass $M_P^{-1}=(8\pi
G_N)^{1/2}$, $\gamma$ is a coupling constant and
$f(\sigma)=\lambda-\hat{\delta} \xi(\sigma)$: the coupling
$\lambda$ may be related to string coupling $g_s$ via $\lambda\sim
1/g_s^2$. The numerical coefficient $\hat{\delta}$ typically
depends on the massless spectrum of every particular
model~\cite{Antoniadis93a}. $V(\sigma)$ is
phenomenologically-motivated field potential. Our approach is a
new take on a familiar system: the Gauss-Bonnet combination is
multiplied by a function of the scalar field. In four dimensions
($d=4$) the GB term makes no contribution if
$f(\sigma)=\text{const}$. It is natural to consider $f(\sigma)$ as
a dynamical variable. This follows, for example, from the one-loop
corrected string effective action~\cite{Narain92a} when going from
the string frame to the Einstein frame to appropriately describe
the universe we observe today, where the equivalence principle is
well preserved and the Newton's constant $G_N$ is (almost)
time-independent. A discussion on cosmological advantages of
modified Gauss-Bonnet theory may be found in~\cite{Nojiri05c}. A
study has been recently made in~\cite{Sami:2005zc} by introducing
higher order (quartic) curvature corrections, but in a background
of fixed modulus field $\sigma$.

The most pertinent question that we would like to ask is: what new
features would a dynamical GB coupling introduce, and how can they
influence a cosmological solution? In this Letter we present an
exact cosmological solution that explains a dynamical relaxation
of vacuum energy to a small value (exponentially close to zero)
after a sufficient number of e-folds, leading to a small deviation
from the $w=-1$ prediction of non-evolving dark energy.

Let us consider the four-dimensional spacetime metric in standard
Friedmann-Robertson-Walker (FRW) form:
\begin{equation}
ds^2=-dt^2 + a(t)^2 \sum_{i=1}^3 (dx^i)^2,
\end{equation}
where $a(t)$ is the scale factor of the universe. In terms of the
following dimensionless variables
\begin{eqnarray}\label{def-variables}
&& x= (\gamma\kappa^2/2) ({\dot\sigma}/{H})^2, \qquad y= \kappa^2
({V(\sigma)}/{H^2}), \nonumber
\\
&& u=8\kappa^2 f(\sigma) H^2, \qquad \quad \ \ \, h=
{\dot{H}}/{H^2},
\end{eqnarray}
the Einstein field equations obtained by varying the metric
$g_{\mu\nu}$ are given by
\begin{eqnarray}
0&=&- {3}+x+y -3 (u^\prime-2 h u),\label{main1} \\
0&=& u^{\prime\prime}+(2-h) u^\prime- 2 h^\prime u -2(2+h)
u h\nonumber\\
&{}&\qquad +\,2h+3+x-y, \label{main2}
\end{eqnarray}
where $X^\prime = \frac{\dd X}{\dd \eta}=\frac{1}{H}\frac{\dd
X}{\dd{t}}= a \frac{\dd {X}}{\dd a}$, $\eta=\int H\, {\dd t}=\ln
(a/a_0)$ is the number of e-folds and $H=\frac{\dot{a}}{a}$ is the
Hubble expansion parameter. The time evolution of $\sigma$ is
given by
\begin{equation}\label{evo-eqn}
\frac{d}{dt}\left({\gamma}\,\dot{\sigma}^2+2\Lambda(\sigma)
\right) =- 6 H \left(\gamma\dot{\sigma}^2\right)-2\delta,
\end{equation}
where $\Lambda(\sigma)\equiv V(\sigma)-f(\sigma){\cal G}$ and
$\delta \equiv f(\sigma)\frac{dG}{dt}$, with ${\cal G}=24
H^2(\dot{H}+H^2)$. We will call $\Lambda(\sigma)$ an effective
potential. Eq.~(\ref{evo-eqn}) may be written as $x^\prime +
2(h+3) x + y^\prime +2 h y - 3 (h+1)(u^\prime-2h u)=0.$

When $f(\sigma)=0$ (or $u=0$), the equations of motion satisfy
simple relationships: $y=3+h$ and $x=- h$. From this we see that
it is enough to introduce only one new parameter (e.g. the
potential $V(\sigma)$) to explain a particular value of slow-roll
type variable $h$ ($\equiv a\ddot{a}/\dot{a}^2-1$) or the
acceleration parameter $\ddot{a}$; if more than one parameter is
introduced in the starting action, as such the case with
$f(\sigma)\neq 0$, then one complicates the problem. This remark
though valid to some extend may have less significance for at
least two reasons. Firstly, a non-trivial coupling between the
field $\sigma$ and the curvature squared terms naturally exists in
one-loop corrected string effective action~\cite{Antoniadis93a},
as well as in the best motivated scalar-tensor theories which
respect most of GR's symmetries, see for
example~\cite{Damour:1994ya,Fabris:2000gz}. Secondly, the
knowledge about the modulus-dependent GB coupling may help one to
construct a cosmological model where $h$ (or the dark energy
equation of state parameter, $w$) is dynamical.

While there can be a myriad of scalar field models with different
forms of $V(\sigma)$ and $f(\sigma)$, inspired by particle physics
beyond the standard model or string theory, not all scalar
potentials may be used to describe the universe that we observe
today. In order for the model to work a scalar field should relax
its potential energy after inflation down to a sufficiently low
value, presumably very close to the observed value of the dark
energy in order to solve the cosmological constant problem. In
this Letter, rather than picking up particular functional forms
for $V(\sigma)$ and $f(\sigma)$, for example, as in
Ref.~\cite{Nojiri05b}, we would like to obtain an exact
cosmological solution, which respects the symmetry of the field
equations.

By solving the field equations (\ref{main1})-(\ref{evo-eqn}), we
can write $x$ and $ y$ (and hence $\Lambda(\sigma)$) as a second
order differential equation in $u$. Notably, $\kappa^2
\Lambda(\sigma)=\frac{H^2}{2} \left[ u^{\prime\prime} +\left(5-
h\right)u^\prime -2(8h+h^2+h^\prime+3)u\right] + H^2(3+h)$, which
generally has a solution composed of {\it homogeneous} and {\it
nonhomogeneous} parts. We refer to the solution of
$u^{\prime\prime} +\left(5- h\right)u^\prime
-2(8h+h^2+h^\prime+3)u=0$, or equivalently,
\begin{equation}\label{homo-poten}
\kappa^2 \Lambda(\sigma)= H^2(3+h)= 3H^2+H H^\prime,
\end{equation}
a {\it homogeneous} solution, which is trivially satisfied for
$f(\sigma)=0$. Effectively, for $f(\sigma)\neq 0$, we will pick
$V(\sigma)= f(\sigma) {{\cal G}} + H^2(3+h)/\kappa^2 $, leaving us
with only one arbitrary function, out of $x(\eta)$, $h(\eta)$ or
$u(\eta)$.

What is the advantage of adding a new term, $f(\sigma){\cal G}$,
if one could make an ansatz like (\ref{homo-poten}) with
$f(\sigma)=0$? An interesting feature of our construction is that
while the contributions coming from both the field potential
$V(\sigma)$ and the coupled GB term, $f(\sigma)G$, can be large
separately, the effective potential, $\Lambda(\sigma)$, can be
exponentially close to zero at late times, as it relaxes to a
small value after a sufficiently large number of e-folds of
expansion. Moreover, even if we do not impose {\it a priori} the
constraint $V(\sigma)- f(\sigma) {{\cal G}} = H^2(3+h)/\kappa^2 $
we arrive at a similar expression for various other solutions;
below we will present one such example.

Given a perfect fluid form for the energy momentum tensor of the
field $\sigma$: $T_{00}=\rho_\sigma$, $T_{ii}=p_\sigma a(t)^2$, we
define the equation of state (EOS) parameter
\begin{equation}
w \equiv
\frac{p_\sigma}{\rho_\sigma}=-\frac{2h+3}{3}=\frac{2q-1}{3},
\end{equation}
where $q \equiv -{a\ddot{a}}/\dot{a}^2$ is the deceleration
parameter. We will assume first that we are dealing with a
canonical-scalar ($\gamma>0$), so that $1+w\geq 0$ (or $h\leq 0$).
When $h \simeq 0$, and hence $H\simeq \text{const}=
H_{\text{inf}}$, $\Lambda(\sigma)$ acts as a cosmological constant
term, i.e. $\Lambda(\sigma)\simeq 3 M_p^2 H_{\text{inf}}^2\equiv
\Lambda_{\text{inf}}$. A major difference from the case of a
cosmological constant $\Lambda_0$, for which
$p_\Lambda=-\rho_\Lambda$ and thus $w_\Lambda=-1$, is that the
field is evolving towards an analytic minimum at
$\Lambda(\sigma)\simeq 0$.

The Gauss-Bonnet term, ${\cal G}=24 \frac{{\dot{a}}^2
\ddot{a}}{a^3}$, changes sign between accelerating ($\ddot{a}>0$)
and decelerating ($\ddot{a}<0$) solutions. In our model, the
effective potential $\Lambda(\sigma)$ ($\equiv
V(\sigma)-f(\sigma){\cal G}$) has to be non-negative in order that
the vacuum becomes a possible ground state. Under the condition
(\ref{homo-poten}), it is required that $h\ge -3$ and consequently
the EOS parameter $w \le 1$. This last inequality may be saturated
only by stiff matter for which $w=1$ (and the velocity of sound
approaches the velocity of light). In general, however, $w<1$
holds and hence $\Lambda(\sigma)>0$.

Under the condition (7) alone, the field potential $V(\sigma)$ may
not always be positive. Then one may ask: Is there any condition
for the (semi-)positivity of the potential? Below we will give the
condition for $V(\sigma)\ge 0 $, though it may not be required for
obtaining inflationary type solutions in the present context.
Furthermore the condition (7) fixes only one out of two arbitrary
parameters in the model. The last arbitrary parameter may be fixed
either by allowing one of the field variables to take a fixed (but
arbitrary) value or by making an appropriate ansatz for one of the
field variables. Here we consider two well motivated examples.

\begin{figure}[ht]
\epsfig{figure=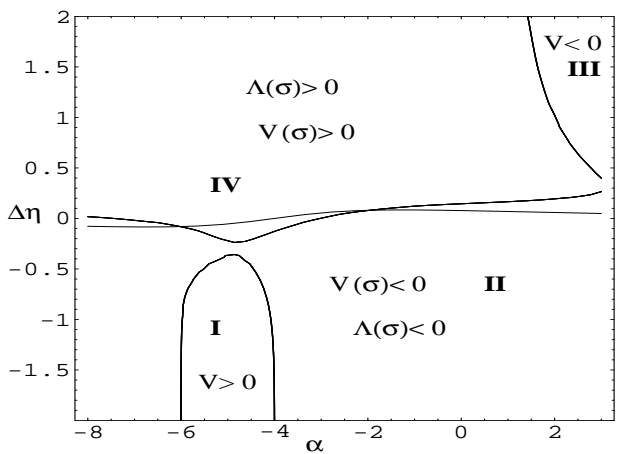,height=2.4in,width=3.2in} \caption{The
$\Delta\eta$ - $\alpha$ phase space, showing regions with
$V(\sigma)>0$ and $V(\sigma)<0$. The almost horizontal line with
$h=-3$ separates the regions between $\Lambda(\sigma)<0$ and
$\Lambda(\sigma)>0$. Here we have set $u_{time}=0.2$. For a large
value of $u_{time}$ ($> 0.2$), the regions I and IV, both with
$V(\sigma)>0$, merge to a single region, while for a small value
of $u_{time}$ ($< 0.2$), the region I moves to a more negative
value of $\Delta\eta$, while the region III moves to a more
positive value of $\Delta\eta$ (and $\alpha$). In any case, the
region with $\Lambda(\sigma)<0$ (or $h+3<0$) is physically less
relevant as it corresponds to the case where the EOS parameter
$w>1$.} \label{figure1}
\end{figure}

If we use the approximation $h\equiv \dot{H}/H^2=H^\prime/H \simeq
\text{const}=h_0$, then, by solving~(\ref{homo-poten}), we find
\begin{equation}\label{sol-u-homo}
H= \e^{\, \int h {\dd \eta}}=H_0 \e^{h_0 \eta}, \quad u
=u\Z1\,\e^{\alpha_- \eta}+u_2\,\e^{\alpha_+ \eta},
\end{equation}
where $H_0$ and $u_i$ are integration constants, and
\begin{equation}
\alpha_\pm = \frac{1}{2}\left(h_0-5 \pm \sqrt{9 h_0^2+54
h_0+49}\right).
\end{equation}
It is the coupling $\alpha_+$ ($\alpha_-$) which is of greater
interest for $\eta\gtrsim 0$ ($\eta\lesssim 0$). The scale factor
is given by $a(t)= a_0 (1- c_0 h_0 t)^{-1/h_0}$, where $c_0
h_0<0$, implying that the universe accelerates in the range
\begin{equation}
-1< h_0 < 0\, \Rightarrow\,  0 >q>-1.
\end{equation}
We then wish to consider the case where $h$ is dynamical. To this
aim, we make no assumptions about the form of $h$, and instead
consider the ansatz that during a given epoch we may make the
approximation \beq\label{sol-u-homo1} u\equiv 8\kappa^2 f(\sigma)
H^2 \approx u_{time} e^{\alpha_{time} \eta} \eeq where the $time$
can be $early$ or $late$, where $|u_{early}|>|u_{late}|$, and
$\alpha_{early} < \alpha_{late}$. This {\it ansatz} is well
motivated, since in the context of string theory the coupling
$f(\sigma)$ is generally a sum of exponential terms (as functions
of $\sigma$ or the number of e-folds) and so is the Hubble
parameter. Interestingly, the ansatz (\ref{sol-u-homo1}) for the
form of $u$ will allow us to write $h$ in a closed form:
\begin{eqnarray}
H&=& H_0\,\cosh\beta (\eta-\eta\Z1)\,\e^{-\hat{\beta}\eta},
\label{main-Hubble}
\\
h&=& -\hat{\beta}+\beta \tanh\beta (\eta-\eta\Z1), \label{main-h}
\end{eqnarray}
where $\eta\Z1$ is a free parameter,
\begin{equation}
\hat{\beta} \equiv 4+\frac{\alpha}{4},\quad \beta \equiv
\frac{1}{4}\sqrt{9\alpha^2+72 \alpha +208},
\end{equation}
and $\alpha=\alpha_{\text{early}}$ or $\alpha_{\text{late}}$.
Writing the expressions for $x(\eta)$ and $y(\eta)$ is
straightforward. Analogous to an inflationary type solution
induced by a conformal-anomaly~\cite{Starobinsky:1980te}, the
solutions given above are singularity-free. Note, for $\Delta\eta
\equiv \eta-\eta\Z1\gtrsim 0$, we always have $\Lambda(\sigma)>0$
for $\alpha\le 1$ (cf Fig.~\ref{figure1}).

Next, consider that $u(\sigma)=8\kappa^2 f(\sigma) H^2 \simeq {\rm
const} \equiv 1/\hat{\alpha}$ and $x \equiv
\frac{(\gamma/2)\dot{\sigma}^2}{H^2}\kappa^2 \equiv x_0$, where
$|\hat{\alpha}|>1$ and $0<x_0<1$, which are reasonably good
approximations at low energy. In this case, $\eta=\pm
\sqrt{{\gamma}/{2x\Z0}}\,(\sigma/M_{Pl})+{\rm const}$. The
solution is again given by (\ref{main-Hubble})-(\ref{main-h}) but
now
\begin{eqnarray}
\hat{\beta}=-\frac{1+\hat{\alpha}}{2}, \quad \beta=\frac{1}{2}
\sqrt{(1+\hat{\alpha})^2+4x_0\hat{\alpha}}.
\end{eqnarray}
In particular, for $x_0\ll 1/(4\hat{\alpha})$, or $\beta\simeq
-\hat{\beta}$, one has $H\propto H_0 [ 1+
\e^{2\beta(\eta-\eta\Z1)}]$; the Hubble expansion rate decreases
with $\eta$ when $\beta(\eta-\eta\Z1)< 0$. The scalar potential
and the coupled GB term may be given by
\begin{eqnarray}
V(\sigma)&=& M_P^2 H^2 \left(-\frac{3}{\hat{\alpha}}
- \frac{6\beta}{\hat{\alpha}}\tanh\beta\tilde{\sigma} -x_0\right),\\
f(\sigma) {\cal G} &=& M_P^2 H^2 \left(
\frac{3+\hat{\alpha}}{16\hat{\alpha}}+
\frac{\beta}{8\hat{\alpha}}\tanh\beta\tilde{\sigma} \right),
\end{eqnarray}
where $\tilde{\sigma}= \sqrt{\gamma/2x\Z0}\,\kappa
(\sigma-\sigma\Z1)$. Note that here $\sigma$ (as well as
$\tilde{\sigma}$) can take a negative value, so that $V(\sigma)>0$
for $3+x_0\hat{\alpha}<6\beta$, or, more precisely, for
$h<(3-x_0)\hat{\alpha}/6$. Initially, $V(\sigma)-f(\sigma) {\cal
G}\neq M_P^2 H^2 (3+h)$, but after a certain number of e-folds,
namely $|\beta(\eta-\eta\Z1)|>2$, so that
$|\tanh\beta\tilde{\sigma}|\to 1$, the relation as above (or
Eq.~(\ref{homo-poten})) may be attained by satisfying
$(1+8x\Z0\hat{\alpha})=(49+8\hat{\alpha})(\beta-\hat{\beta})$.
That is, the condition like (\ref{homo-poten}) puts an additional
constraint on the model parameters but it may not exhaust the
basic characteristics of the model.

A question may be raised as to whether inflation is due to the new
term in the action, $f(\sigma){\cal G}$, or simply due to the
potential $V(\sigma)$. In string theory context, the coupling
$f(\sigma)$ may be expanded as $f(\sigma)\propto
\left[\frac{2\pi}{3} \cosh(\sigma)-\ln(2)\right]+{\rm const}$. At
early times, $\sigma<0$, the potential $V(\sigma)$ is expected to
dominate the Gauss-Bonnet contribution, $f(\sigma) {\cal G}$. At
the start of inflation, ${\cal G}=24
\left(\frac{\dot{a}}{a}\right)^2 \frac{\ddot{a}}{a}\to 0$.
Inflation is mainly due to the potential which is not essentially
flat; in such a case the Hubble expansion rate could naturally
drop to a significantly low value after a sufficient number of
e-folds. This can be understood also from some earlier studies on
the subject~\cite{Antoniadis93a,Easther:1996yd}. With
$V(\sigma)=0$, the period of acceleration is short and the number
of e-folds is order of unity; a model with vanishing potential may
not be suitable for early inflation. At late times, the potential
drops relatively faster as compared to the coupled Gauss-Bonnet
term, so the contribution of the potential nearly equates or
slightly exceeds the contribution of the GB term.

\begin{figure}[ht]
\epsfig{figure=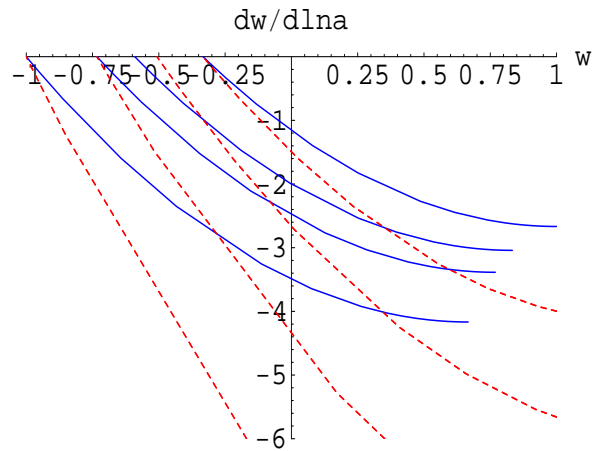,height=2.4in,width=3.3in} \caption{The
$w - w^\prime$ phase space occupied by the scalar field, in the
range $0\geq \eta-\eta\Z1\geq 5$. From left to right
$\alpha=-6,-5.38, -5,-4$ (solid lines) and $\alpha=+1, -0.0148,-1,
-2$ (dashed lines). The universe is not accelerating for
$-4<\alpha<-2$. The variation of $w^\prime$ w.r.t. $w$ is
significant in the range $|\eta-\eta\Z1|\lesssim 2.5/\beta$. This
happens when the red-shift factor $z < 0.8~(2.49)$, since
$\beta<4.25~(>2)$, if $|\eta-\eta\Z1|$ is to be related to $z$ via
$1+z=\e^{\eta-\eta\Z1}$. For $|\eta-\eta\Z1|\gtrsim 2.5/\beta$,
the field is almost frozen, $w\simeq \text{const}$; $w\leq -1/3$
if $\eta>\eta\Z1$, while $w>0$ if $\eta<\eta\Z1$.} \label{figure2}
\end{figure}

\begin{figure}[ht]
\epsfig{figure=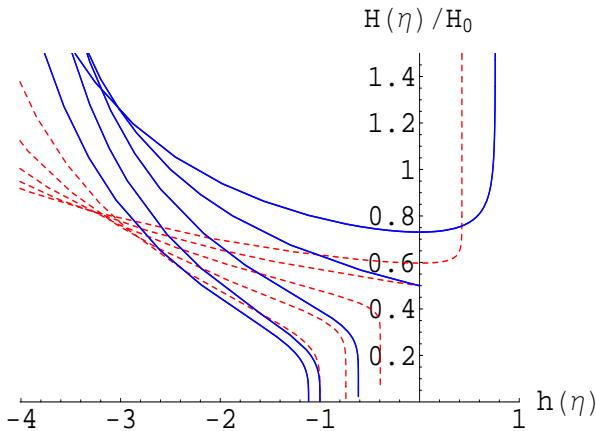,height=2.3in,width=3.3in} \caption{ The
$h - H$ phase space occupied by the scalar field, in the range
$-5\leq \eta-\eta\Z1\leq 5$. From top to bottom $\alpha=-7,
-6,-5,-4,-3$ (solid lines) and $\alpha=2,1, 0,-1,-2$ (dashed
lines). For a canonical scalar (i.e. $\gamma>0$), $\alpha$ is only
well defined for $-6< \alpha<1$. The Hubble parameter decreases
with proper time for the coupling $-6 < \alpha < 1$, while it
increases with proper time for the coupling $\alpha<-6$ or
$\alpha>1$, heading to a value $w<-1$, since $h>0$. In this last
case, we get a negative kinetic energy for the field $\sigma$, in
some regions of field space, which then requires $\gamma<0$
(phantom field).} \label{figure3}
\end{figure}

Let us discuss the result (\ref{main-Hubble})-(\ref{main-h}) in
some details. The universe starts to expand as $\eta$ increases
from $\eta_i$, but it accelerates only when $\eta\gtrsim \eta\Z1$.
Eventually when $\eta \gtrsim \eta\Z1+2.5/\beta$, the scalar field
begins to freeze in, so that $w\leq -1/3$; the actual value of $w$
depends on the value of $\alpha$ ($=\alpha_{\text{early}}$), see
Figs.~\ref{figure2} and \ref{figure3}. After a certain number of
e-folds, say $N$, our approximation, (\ref{sol-u-homo1}), with the
$time$ being $early$ breaks down. Sometime later, subsequent
evolution will be controlled by (\ref{sol-u-homo1}) with the
$time$ being $late$. For $\eta \lesssim \eta_{late}$ the universe
is in a deceleration phase which implies that inflation must have
stopped during the intermediate epoch. As $\eta$ crosses
$\eta_{late}$, the universe begins to accelerate for the second
time.

A large shift in Hubble rate may be required to explain relaxation
of vacuum energy from a natural value $\Lambda(\sigma)\sim M_P^2
H_{b}^2\sim {10}^{-8} M_P^4 $ ({\it before} inflation) to a
sufficiently low value $\Lambda(\sigma)\sim M_P^2 H_{a}^2 \sim
{10}^{-120} M_P^4$ ({\it after} inflation), viz $H_{b} \sim
{10}^{23}\,\text{eV}$ and $H_{a} \sim {10}^{-33}\,{\text eV}$. One
thus calculates the change in $H$ during an inflationary epoch of
$N$ e-folds, by considering the ratio
\begin{equation}
\varepsilon \equiv \frac{H (\eta_i+N)}{H (\eta_i)} = \frac{\cosh
\beta (\Delta\eta+N)\,\e^{-\hat{\beta}
N}}{\cosh\beta(\Delta\eta)},
\end{equation}
where $\Delta\eta\equiv \eta_i-\eta\Z1$, with $\eta_i$ being the
time at on-set of inflation. In our model, the total number of
e-folds $N$ required to get a small ratio, like $\varepsilon \sim
{10}^{-56}$, depends on the value of $\alpha$, which is related to
the slope of the potential coming purely from the Gauss-Bonnet
coupling, $f(\sigma)\,{{\cal G}}$. The value of $N$ would be
small, $N \sim 129$, for $\alpha \gtrsim -2$ (or $\alpha \lesssim
-4$), while it would be large for $\alpha\lesssim 1$ (or
$\alpha\gtrsim -6$), $N \sim {\cal O}(300)$.

The solution (\ref{sol-u-homo})-(\ref{main-h}) may also be used to
explain a cosmic acceleration of the universe at late times, i.e.,
at an energy scale many orders of magnitude lower from inflation.
The present value of the parameter $h$ must be determined by
observation. For $|\eta-\eta\Z1|\gtrsim 2.5/\beta$, $h(\eta)$
varies with the logarithmic time, $\eta$, almost negligibly. The
equation of state parameter for the dark energy is $w=-(2h+3)/3$.
From this, and using (\ref{main-h}), we get $w\geq -1$ and
$\hat{\beta}\geq \beta$ for $-6\leq \alpha \leq 1$, so that $H$
decreases with proper time (Fig.~\ref{figure3}). For instance, if
the observed value of the deceleration parameter corresponds to
\begin{equation}
q_{\text{obs}}=-1-h_{\text{obs}} \simeq -\hat{\beta}-\beta\simeq
-0.6,
\end{equation}
then $\alpha=-5.3851$ or $-0.0148$ and hence $w\simeq -0.73$. One
may achieve $w<-1$, for $\eta\gg \eta\Z1+2.5/\beta$, by allowing
$\alpha<-6$ or $\alpha>1$. In this case, however, since
$\beta>\hat{\beta}$, $H$ increases with proper time. There are
arguments that the present cosmological data may even favor a
value $w<-1$\cite{Caldwell:03vq}; if this is the case, in our
model, one has to allow $\alpha>1$ or $\alpha<-6$.

The equation of state parameter $w$ changes significantly with the
exponent $\alpha$ but not much with $\eta$, except in the range
$|\eta-\eta\Z1|<2.5/\beta$. That is, the scalar field comes to
dominate the universe, presumably for the second time, when
$0<\eta-\eta_{\text{late}}\lesssim 2.5/\beta$, and it gradually
becomes a constant value, $w\simeq \text{const}$, when $\eta>
\eta_{\text{late}}+2.5/\beta$. At late times, one finds
$-1\lesssim w\leq -1/3$, depending upon the value of $\alpha$, see
Fig.~\ref{figure3}. Whether the scalar field $\sigma$ dominates
the future evolution of the universe depends on energy density of
matter system.

In a known form for the GB coupling, derived from the heterotic
string theory, $\xi(\sigma) \sim \ln
2-\frac{2\pi}{3}\cosh\sigma$~\cite{Antoniadis93a,Easther:1996yd}.
It follows that $f(\sigma) \sim c_0 + c_1\left (\e^{\sigma}+
\e^{-\sigma}\right)$. For $\sigma>0$, only the first two terms are
relevant. To this end, we slightly modify the ansatz for
$u(\sigma)$, namely
$u(\sigma)\equiv u_0(1+ u\Z1\,\e^{\alpha \eta})$
with $\alpha\eta<0$, which may be appropriate at some intermediate
epoch. Note that $R/{2\kappa^2}=3(2H^2+\dot{H})/\kappa^2$ and
$f(\sigma) {\cal G}=3u(H^2+\dot{H})/\kappa^2$. We demand $0<
|u|\ll 2$, so that the GB contribution to the action is only
sub-dominant. The solution for $h$ is modified as
\begin{equation}
h=-(\hat{\beta}+\beta)\left[1+\left(\frac{4+\sqrt{13}}
{\hat{\beta}+\beta}-1\right) \sqrt{1+u\Z1\e^{\alpha\eta}}\,
K(\eta)\right]
\end{equation}
where $K(\eta)\equiv (L_{P_+}+ c L_{Q_+})/(L_{P_-}+ c L_{Q_-})$
with $L_{P_{\pm}}$ ($L_{Q_{\pm}}$) being associated Legendre
functions of first (second) kind,
$L_{P(Q)}\left(\frac{2\beta}{\alpha} \pm \frac{1}{2},
\frac{2\sqrt{13}}{\alpha},\sqrt{1+u\Z1\e^{\alpha\eta}}\right)$,
and $c$ an integration constant. For $|\eta-\eta\Z1|\gg 0$ and
$\alpha<0$, so $u(\sigma) \to u_0$, we find $h\simeq -4+\sqrt{13}
\tanh \sqrt{13}(\eta-\eta\Z1)\simeq -0.3944$ and hence
$w_{DE}\simeq -0.7370$. This value of $w$ is realized only at the
asymptotic future, where $u(\sigma)$ attains a small (constant)
value, while its present value, $w_{DE}^{0}$, may be less than
$w_{DE}$. In the presence of matter coupling, $\Omega_m\simeq
0.3$, one defines an effective EOS parameter, $w_{eff}$, whose
value can be greater than $w_{DE}^{0}$.

It is generally valid that the transformation from the string
frame to the Einstein frame introduces couplings between standard
model fields and massless degrees of freedom, such as, dilaton
$\phi$ (a spin-$0$ partner of spin-$2$
graviton)~\cite{Damour:2002mi}. Simple arguments such as the
absence of ghost, thereby guaranteeing the stability of the field
theory, would suffice to rule out a wide class of scalar-tensor
models. In our model, we have assumed, rather implicitly, that
dilaton is either constant or it is extremized,
$\frac{\dd\lambda(\phi)}{\dd\phi}\simeq 0$, so the effects coming
from terms like $\lambda(\phi) {\cal G}$ are negligible. Only the
dynamical field in the model is the modulus $\sigma$, which is
expected to have a non-zero mass, i.e., a non-zero vacuum
expectation value. In string theory context~\cite{Antoniadis93a},
there is no coupling between the field $\sigma$ and the
Einstein-Hilbert term. And there is no direct coupling of the
field $\sigma$ to matter, in view of precise verifications of the
weak equivalence principle. Nevertheless, it is useful to assume
that the running of the field $\sigma$ is negligibly small at late
times, namely~\cite{Farese:2004cc,Amendola:2005}
\begin{equation} H^2 \dot{\sigma}
\frac{df(\sigma)}{\dd\sigma} < {10}^{-3},
\end{equation}
so as to avoid a possible conflict with the time-variation of
Newton's constant under Newtonian approximation.

In conclusion, we find that the dark energy hypothesis fits into a
low energy gravitational action where a scalar field is coupled to
the curvature squared terms in the Gauss-Bonnet combination. It is
established that a GB-scalar coupling can play an important and
interesting role in explaining both the early and late-time
acceleration of the universe, with singularity-free solutions. An
important feature of our solutions is also that the effective
potential (or dark energy) can naturally relax to a sufficiently
low value, $\Lambda\sim {10}^{-120} M_P^4$, after $> 125$ e-folds
of expansion.

It is shown that one may cross the cosmological constant/phantom
barrier, $w=-1$, only by allowing the parameter $\alpha$, coming
from the GB coupling, such that the Hubble rate grows with proper
time, or by allowing a non-canonical scalar, $\gamma<0$ (phantom
field).

A number of follow-up studies may be devised including: a fit to
determine how well the potential may be tuned to the present value
of the cosmological constant, and how well slow roll parameters
fit the spectral index in an imprint on the cosmic microwave
background anisotropy and mass power spectrum. Some of the results
appear elsewhere~\cite{Ish05d}. It is possible to generalize the
model studied here to a system of two scalar fields, by
introducing a dilaton field $\phi$~\cite{Ish06a}.

\medskip
{\bf Acknowledgement}\ This work was supported in part by the
Marsden fund of the Royal Society of New Zealand.

\end{document}